# Application of Convolutional Neural Network to TSOM Images for Classification of 6 nm Node Patterned Defects

**Ravikiran Attota**
Frederick, MD, USA
attota@outlook.com

**Abstract:** *With the rapid growth in the semiconductor industry, it is becoming critical to detect and classify increasingly smaller patterned defects. Recently machine learning, including deep learning, has come to aid in this endeavor in a big way. However, the literature shows that it is challenging to successfully classify defect types at the 6 nm node with 100% accuracy using low-cost and high-volume-manufacturing compatible optical imaging methods. Here we combine a convolutional neural network (CNN) with that of an optical imaging method called through-focus scanning optical microscopy (TSOM) to successfully classify patterned defects for the 6 nm node targets using simulated optical images at the 193 nm illumination wavelength. We demonstrate the successful classification of eight variations of the defects, including the 3 nm difference in the defect size in one dimension, which is over 50 times smaller than the illumination wavelength used.*

## 1. Introduction

Over the years, significant developments in defect detection using optical methods have been reported [1–10]. The rapid pace of developments in the semiconductor industry also brings the challenge of detecting increasingly smaller defects [11,12]. To improve the final yield, it is vital to minimize defects. Defect detection and proper classification are critical to successfully detect and minimize certain types of defects referred to as "killer defects" in patterned wafers [13–16,12]. The application of optical methods for this purpose is advantageous as they are generally considered to have low cost, high throughput, and non-contact. Because of the noncontact nature, the measurement process itself does not induce further contamination.

In this challenging environment, recently machine learning (ML) methods have been applied to assist in improving defect detection [12–22]. So far, the application of the ML methods appears to improve the defect detection process [14,23]. The authors of several papers have presented the mechanics of the ML, especially using the convolutional neural network (CNN) method for defect detection [13,20,24–26]. For image-based tools, the noise seems to be the limiting factor [13,14]. At the 9 nm node, the authors accurately classified the shapes and estimated the sizes of defects [14]. This was further improved for the smaller 6 nm node defects using 193 nm illumination [13]. However, at the 6 nm node pattern, the authors reported accurate classification of three out of the four defect types studied using the best-performing CNN method. Here, we report further improvement in the classification accuracy of defects at the 6 nm node by combining CNN with the through-focus scanning optical microscopy (TSOM) imaging method [27–37].

In TSOM, the imaging target is scanned through the focus simultaneously collecting the out-of-focus optical images. Stacking the set of through-focus images thus obtained at their respective focus positions results in three-dimensional (3D) optical data. Extracting a vertical cross-section through this 3D data results in a TSOM image. A TSOM image plane is perpendicular to a conventional optical image plane [38]. A TSOM image contains information not only at the best focus position optical image, but also out-of-focus images, and thus it contains substantially additional optical information regarding the target being imaged. The TSOM image intensities are typically normalized before further processing [39]. A differential TSOM image is a pixel-by-pixel difference obtained using two TSOM images collected from two slightly varying targets and has some unique characteristics beneficial for 3D shape metrology [36]. TSOM has been applied for 3D shape metrology of target sizes ranging from sub-nanometer to over 100 $\mu m$, with sub-nanometer measurement sensitivity in all dimensions [33]. Some high throughput through-focus image data collection methods are discussed here [40]. Several authors applied ML techniques to TSOM which improved semiconductor metrology [24–26,41,42]. Here we hypothesize that this additional optical information available in TSOM images perhaps helps in improving defect classification.

## 2. Simulation details

For this work, we chose a 6 nm node pattern. In Fig. 1 we show the three variations of a bridging defect and five variations of a line extension defect, for a total of 8 variations of the defects studied here. We chose a set of defect types similar to the ones presented in [13], as these are considered harder to detect. We have added a few more variations to this set to expand this study. We adopted a similar naming convention as defined by SEMATECH [23]. The depth of the pattern is 15 nm. The dimensions of the selected defects are shown in Table 1. The smallest difference in the defect size in one dimension is 3 nm between the By and Cy defects in the horizontal direction.

We performed optical simulations using a commercial program that implements a finite-difference time-domain method to model electromagnetic field scattering [43]. We used an illumination wavelength of 193 nm, illumination numerical aperture of 0.12, collection numerical aperture of 0.95, pixel size at the sample of 10 nm x 10 nm, focus step size of 50 nm, and a through-focus range of 4 $\mu m$. Images were obtained at two polarizations for all the targets: zero degrees where the electric field is in the horizontal orientation, i.e., perpendicular to the lines, and 90° where the electric field is in the vertical orientation, i.e., along the lines. Here we aim to theoretically evaluate the efficacy of combining TSOM with CNN for defect classification. For this reason, a realistic line edge roughness was not implemented during this study. However, we added noise to the simulated images.

We collected TSOM images with their x-axis perpendicular to the lines and passing through the defects. We show typical TSOM images as obtained from the simulations in Figs. 2(a) and 2(b) for the two polarizations for the Bx2 defect type, using a pseudo color scheme. The range of optical intensity present in a TSOM image is represented by a color scale bar on the right side of the TSOM images. The Bx2 defect shows the highest optical signal among the selected defects, as defined by the optical intensity range (OIR). The OIR is defined as the maximum intensity range present in a TSOM or a differential TSOM image and multiplied by 100 [29]. The defect type CxS produced the smallest signal as shown in Fig. 3(a) and 3(b).

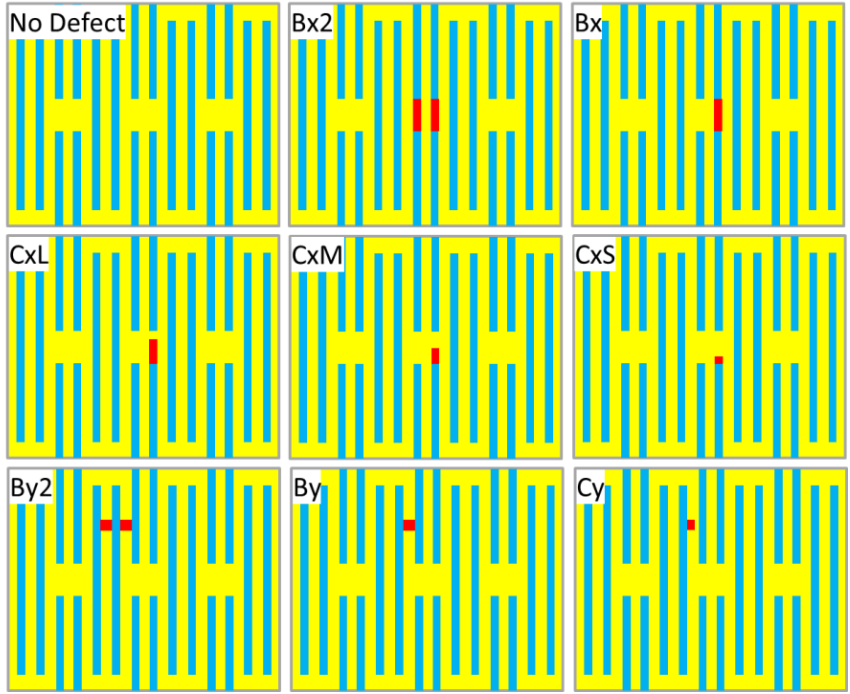

Figure 1. The eight defect variations selected are shown in red color along with the ideal no-defect target. The naming convention is shown in the inset. The substrate represented in yellow color is of crystalline silicon, and the pattern represented in cyan color is of amorphous silicon. The critical dimension of the pattern is 6 nm.

Table 1. Dimensions of the selected defects. The smallest difference in the defect size in one dimension is 3 nm between By and Cy in the horizontal direction.

| S.No. | Defect type | Dimensions, nm (Horizontal x Vertical) | Max. intensity polarization | Additional detail |
|---|---|---|---|---|
| 1 | Bx2 | 6 x 24 | 90 | Two defects |
| 2 | Bx | 6 x 24 | 90 | Single defect |
| 3 | CxL | 6 x 18 | 90 | Single defect |
| 4 | CxM | 6 x 12 | 90 | Single defect |
| 5 | CxS | 6 x 6 | 90 | Single defect |
| 6 | By2 | 9 x 9 | 0 | Two defects |
| 7 | By | 9 x 9 | 0 | Single defect |
| 8 | Cy | 6 x 9 | 0 | Single defect |

In previous publications, we reported that experimental noise has an OIR of less than 1 [44]. For this study, we chose a noise OIR of 0.8 and added random noise to the TSOM images obtained from the simulations. We show the noise-added TSOM images in Figs. 2(a1) and 2(b1) for the Bx2 defect, and in Figs. 3(a1) and 3(b1) for the CxS defect. The addition of noise typically increases the OIR. Due to the lower signal present for the CxS defect, adding the same amount of random noise affects the appearance of the TSOM image substantially (Figs. 3(a1) and 3(b1)). From Figs. 2 and 3 we observe that 90° polarization illumination results in a stronger optical signal for the line-extension type of defects. In this initial theoretical evaluation of the efficacy of combining TSOM with CNN for the defect classification, we used only the polarization that produced the highest OIR as given in Table 1. We also assumed the highest photon densities in this study.

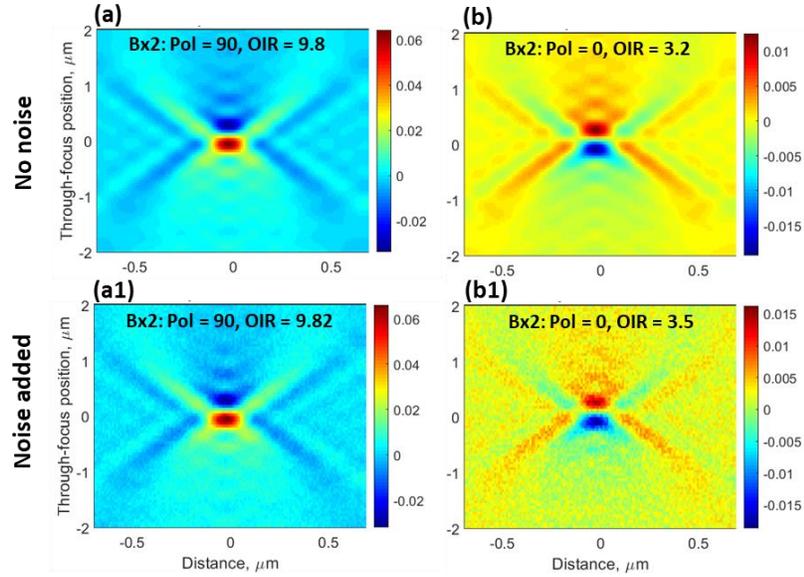

Figure 2. TSOM images obtained from the optical simulations for the Bx2 defect at (a) 90° and (b) 0° polarization illuminations. Random noise added TSOM images at (a1) 90° and (b1) 0° polarization illuminations. The color scale is set to automatic.

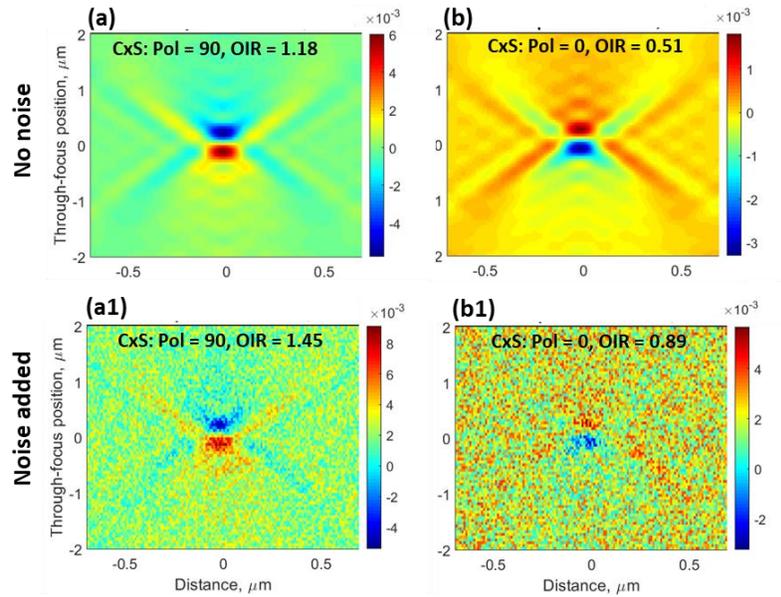

Figure 3. TSOM images obtained from the optical simulations for the CxS defect at (a) 90° and (b) 0° polarization illuminations. Random noise added TSOM images at (a1) 90° and (b1) 0° polarization illuminations. The color scale is set to automatic.

In conventional defect analysis using optical microscopes, a single optical image is collected at a given fixed focus position, and then defect analysis is done on that optical image. However, this single optical image may not capture the highest optical signal from the defect. This is because different types of defects exhibit the highest signal at different focal positions [7]. A plot of focus metric with focus position [44] shows the focus position where the optical signal is maximum. In Fig. 4 we show the focus metric plots for the eight variations of the selected defects. The red vertical line represents the fixed focal position used by conventional methods. This figure shows that the focal position where the optical signal from the defect is maximum varies for each defect type. Hence no single fixed focal position image can capture the maximum signal from the defects for all the defect types and variations, thus reducing the defect signal and analysis efficiency. However, a TSOM image due to its inherent through-focus image acquisition method captures all the highest defect intensities, thus enabling higher efficient defect analysis.

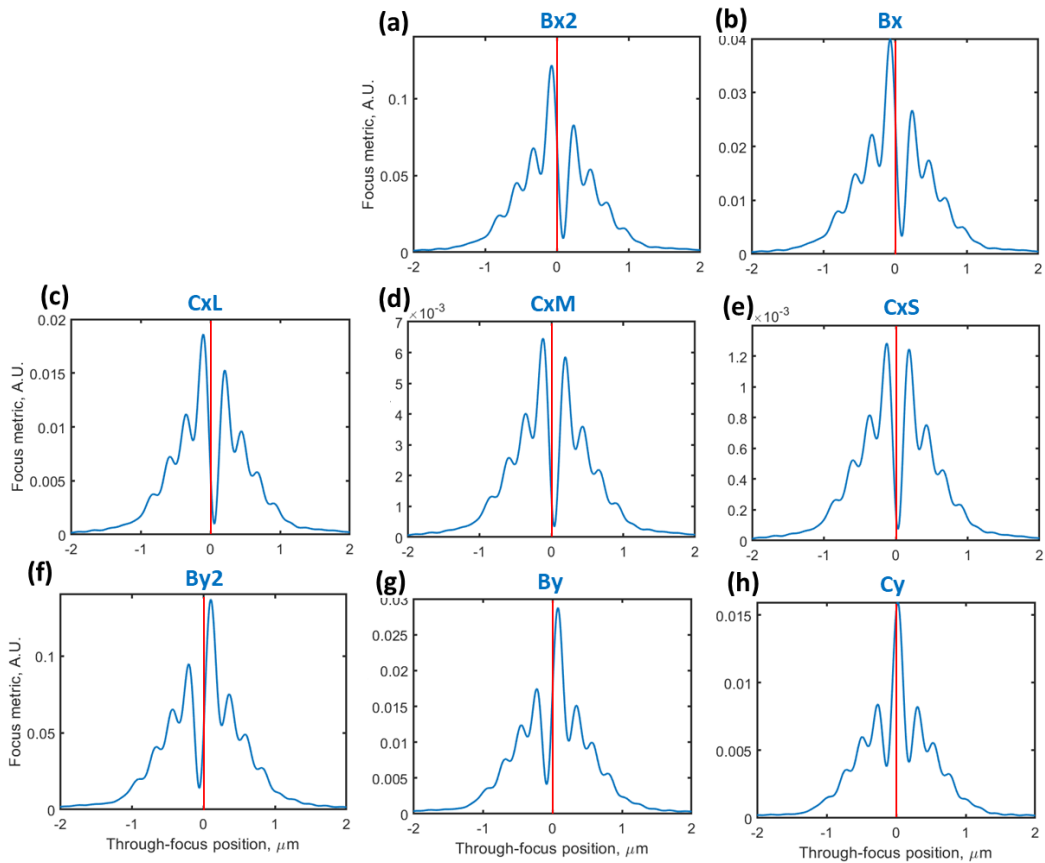

Figure 4. Plots of the focus metric with the focus position for the eight variations of the defects selected. The red vertical lines represent conventionally used fixed focal positions. The highest signal from the defect is available at the focal position where the focus metric is the maximum.

## 3. CNN analysis

For image-based ML analysis, CNN produces the best result [13], and hence in this study, we chose to apply only CNN. For this work, we did data munging using MATLAB-like free software Octave, and neural network analysis using TensorFlow. In this supervised learning method, we initially randomized all the image data, out of which we used 80% of the data for the training purpose and the rest for testing. For the CNN analysis, we used a single channel TSOM image size of 61 x 81 pixels, 32 filters, a kernel size of 3 x 3, a max pooling size of 2 x 2, ReLU for activation, and Adam for optimization.

The CNN model accuracy quickly reached 100% after 5 Epochs, and the loss function reduced to 0.008 after 10 Epochs as shown in Figs. 5(a) and 5(b), respectively. The resulting confusion matrix in Fig. 5(c) shows that all the eight variations of the two types of defects studied here were accurately classified using this CNN model, including the four different-sized line extensions, and the 3 nm difference in the bridge type defects.

We believe the reason for the accuracy to reach 100% quickly is due to the reasonably distinct TSOM images for all the eight defect variations to start with. The additional out-of-focus image information contained in the TSOM images perhaps aids in their easy differentiation. Based on this result we predict further room for improvement. Successful defect detection and classification for even smaller defects, possibly for a 4 nm node pattern, appears feasible.

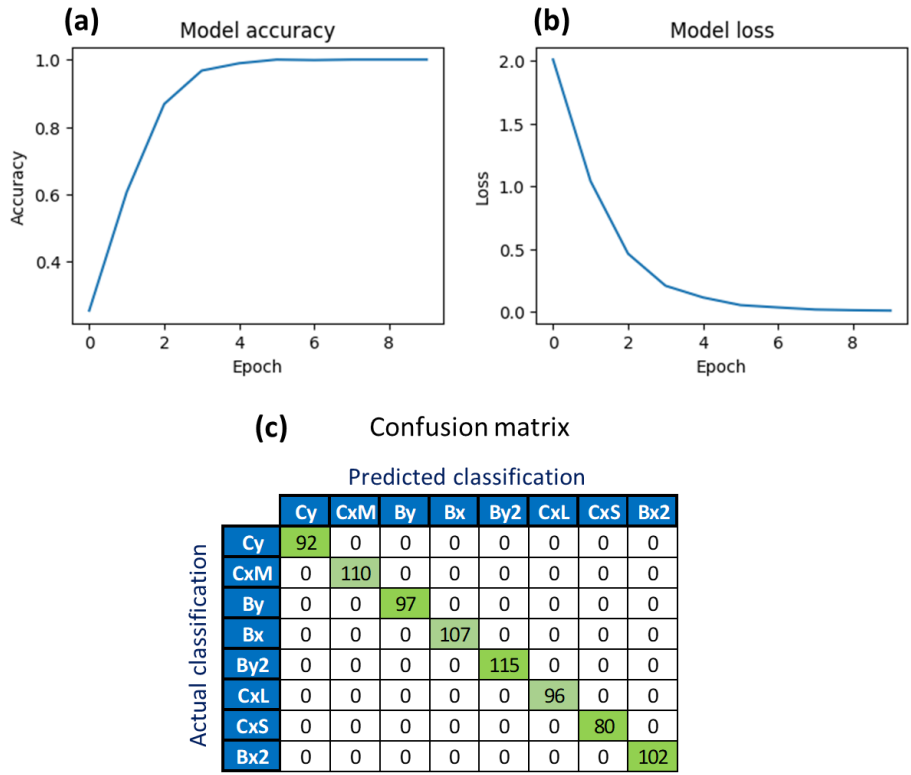

Figure 5. Plots of (a) Model accuracy, and (b) Model loss function with the number of Epochs. (c) The resultant Confusion matrix shows a 100% accurate classification of the defects. The number of Epochs = 10.

To study the progress of the model we obtained the CNN result after three Epochs (Fig. 6) before the model reached 100% accuracy. When the model was not fully optimized, we observed the miss-classification of CxL as CxM, and CxM as CxS. This is reasonable as these defect types are physically closely related. However, as seen in Fig. 5 this miss-classification disappears when the model was fully optimized.

Confusion matrix after 3 Epochs

Predicted classification

| Actual classification | Cy | CxM | By | Bx | By2 | CxL | CxS | Bx2 |
|---|---|---|---|---|---|---|---|---|
| Cy | 92 | 0 | 0 | 0 | 0 | 0 | 0 | 0 |
| CxM | 0 | 64 | 0 | 0 | 0 | 0 | 46 | 0 |
| By | 0 | 0 | 97 | 0 | 0 | 0 | 0 | 0 |
| Bx | 0 | 0 | 0 | 107 | 0 | 0 | 0 | 0 |
| By2 | 0 | 0 | 0 | 0 | 115 | 0 | 0 | 0 |
| CxL | 0 | 16 | 0 | 2 | 0 | 78 | 0 | 0 |
| CxS | 0 | 0 | 0 | 0 | 0 | 0 | 80 | 0 |
| Bx2 | 0 | 0 | 0 | 0 | 0 | 0 | 0 | 102 |

Figure 6. The confusion matrix was obtained after 3 Epochs. The accuracy was about 90% and the loss function had a value of about 0.5.

## 4. Conclusion

Using the optical simulations, we collected the TSOM images for the two defect types with the eight variations for the 6 nm node pattern. The CNN model developed classified all the defects with 100% accuracy, including the 3 nm difference in the defect size in one dimension. This initial study shows that the combination of the TSOM imaging method with that of the CNN method appears promising for defect classification.

## Acknowledgments

The author thanks Tom Pistor of Panoramic Technology Inc., for providing access to an optical simulation program.